\documentclass[a4paper,11pt]{article}
\pdfoutput=1 
\usepackage{jheppub} 
\usepackage{bbm,bm,graphicx,mathtools,color,hyperref,slashed}
\usepackage{amssymb,amsmath}
\usepackage[T1]{fontenc}

\graphicspath{{./Figures/}}

\newcommand{\diff}{\mathrm{d}}
\newcommand{\p}{\partial}

\newcommand{\Diff}{{\mathcal{D}}}

\newcommand{\be}{\begin{equation}}      
\newcommand{\ee}{\end{equation}}      
\newcommand{\bea}{\begin{eqnarray}}      
\newcommand{\eea}{\end{eqnarray}}

\newcommand{\im}{\mathrm{i}}

\newcommand{\calG}{\mathcal{G}}

\newcommand{\calZ}{\mathcal{Z}}

\title{Circle compactification and 't Hooft anomaly}

\author[a]{Yuya Tanizaki,}
\author[b,c,d]{Tatsuhiro Misumi,}
\author[c,d]{Norisuke Sakai}

\affiliation[a]{RIKEN BNL Research Center, Brookhaven National Laboratory, Upton, NY 11973, USA}
\affiliation[b]{Department of Mathematical Science, Akita University, 1-1 Tegata Gakuen-machi, Akita 010-8502, Japan}
\affiliation[c]{Department of Physics, and Research and Education Center for Natural Sciences, Keio University, 4-1-1 Hiyoshi, Yokohama, Kanagawa 223-8521, Japan}
\affiliation[d]{iTHEMS, RIKEN, 2-1 Hirosawa, Wako, Saitama 351-0198, Japan}

\emailAdd{yuya.tanizaki@riken.jp}
\emailAdd{misumi@phys.akita-u.ac.jp}
\emailAdd{norisuke.sakai@gmail.com}

\abstract{
Anomaly matching constrains low-energy physics of strongly-coupled field theories, but it is not useful at finite temperature due to contamination from high-energy states. 
The known exception is an 't Hooft anomaly involving one-form symmetries as in pure $SU(N)$ Yang-Mills theory at $\theta=\pi$. 
Recent development about large-$N$ volume independence, however, gives us a circumstantial evidence that 't Hooft anomalies can also remain under circle compactifications in some theories without one-form symmetries.  
We develop a systematic procedure for deriving an 't Hooft anomaly of the circle-compactified theory starting from the anomaly of the original uncompactified theory without one-form symmetries, where the twisted boundary condition for the compactified direction plays a pivotal role. As an application, we consider $\mathbb{Z}_N$-twisted $\mathbb{C}P^{N-1}$ sigma model and massless $\mathbb{Z}_N$-QCD, and compute their anomalies explicitly. 
}

\begin{document}
\maketitle
\section{Introduction}\label{sec:introduction}

Quantum field theory (QFT) provides us a universal description about collective quantum phenomena that appear in huge varieties of physical systems from particle and nuclear physics to condensed matter physics.  
However, we often encounter the situation where things we observe at low energies look completely different from microscopic degrees of freedom describing QFT. 
That happens when QFTs of our interest is strongly coupled, and nonperturbative aspects of QFTs are still in big mystery. 

One of the possible approaches to tackle this situation is to find a rigorous nature of QFTs, 
especially related to symmetries and topologies. 
Symmetry has always played a key role in the development of QFTs; for instance, the idea of spontaneous symmetry breaking classifies traditional phases of matter following Landau's characterization~\cite{landau1937theory,ginzburg1950theory}. 
In order to refine the data of QFTs related to symmetry, one can try to promote global symmetry to local gauge symmetry, but sometimes topology related to the symmetry gives an obstruction. Such obstruction is called an 't Hooft anomaly, which is of great importance because of its preservation under the renormalization group flow~\cite{tHooft:1979rat, Frishman:1980dq, Coleman:1982yg}: 
An 't Hooft anomaly computed by the low-energy effective theory must be equal to that of microscopic degrees of freedom. It provides us an important consistency check to determine the structure of vacuum and its low-energy excitations and we can use it irrespective of QFTs of our interest being strongly coupled or not. 
Originally, anomaly matching was proposed for studying chiral symmetries of gauge theories with massless fermions~\cite{tHooft:1979rat, Frishman:1980dq, Coleman:1982yg}.  Recent development on topological phases of matter pushes that notion further~\cite{Vishwanath:2012tq, Wang:2014pma, Kapustin:2014lwa, Kapustin:2014zva, Cho:2014jfa} and it is now applicable also for systems with discrete symmetries, higher-form symmetries, and so forth, to derive nontrivial consequences on vacuum structures~\cite{Witten:2015aba,Seiberg:2016rsg,Witten:2016cio, Tachikawa:2016nmo, Gaiotto:2017yup,Wang:2017txt, Tanizaki:2017bam, Komargodski:2017dmc, Komargodski:2017smk, Cho:2017fgz,Shimizu:2017asf, Wang:2017loc, Metlitski:2017fmd,Kikuchi:2017pcp, Gaiotto:2017tne}.

Although anomaly matching is a powerful technique to study nonperturbative physics, it cannot uncover details of dynamical aspects of QFTs and just provides us a consistency condition. 
For example, $SU(N)$ Yang--Mills theory is believed to exhibit confinement in four-dimensional spacetime, and an 't Hooft anomaly can tell us about some additional information on vacuum assuming confinement, but it does not show how confinement can happen. 
Analytic computation of confinement in four dimensions is currently impossible because of its strong coupling nature. Still, it is found that the confinement of $SU(N)$ Yang-Mills theory is realizable on $\mathbb{R}^3\times S^1$ by adding several massive adjoint fermions or deformations of the action itself, and this confinement is calculable with reliable semiclassical computations~\cite{Unsal:2007vu, Kovtun:2007py, Unsal:2007jx, Unsal:2008ch, Shifman:2008ja, Shifman:2009tp}. 
What is more interesting is that this semiclassical confinement is argued to be adiabatically connected to the confinement in the strongly-coupled regime by decompactifying the circle $S^1$ especially in the large-$N$ limit. 
This finding motivated many studies of various asymptotically-free field theories by compactifying one direction to a circle with an appropriate boundary condition, and it is expected to map the strongly-coupled dynamics into the semiclassical regime without losing its essential information~\cite{Cossu:2009sq, Cossu:2013ora, Argyres:2012ka, Argyres:2012vv, Dunne:2012ae, Dunne:2012zk, Poppitz:2012sw, Anber:2013doa, Basar:2013sza, Cherman:2013yfa, Cherman:2014ofa, Misumi:2014raa, Misumi:2014jua, Misumi:2014bsa, Dunne:2016nmc, Misumi:2016fno, Cherman:2016hcd, Fujimori:2016ljw, Sulejmanpasic:2016llc, Yamazaki:2017ulc, Buividovich:2017jea, Aitken:2017ayq}. 

In this paper, we would like to make a connection between these two recent developments of nonperturbative QFTs; adiabatic circle compactification and 't Hooft anomaly matching. 
If the vacuum structures of the original and circle-compactified theories are really adiabatically connected, it is natural to think that both vacuum structures reproduce the same 't Hooft anomaly matching condition. 
However, there is the following difficulty in this idea: Anomaly is renormalization group invariant, and it is matched by the vacuum or its low-energy excitations. 
Since other high-energy states do not produce the anomaly, the effect of anomaly would disappear once those high-energy states give dominant contributions at finite temperature.  
How can this observation be consistent with the story about adiabatic continuity? 
In order to understand the situation better, we consider two quick examples. 

Let us consider a three-dimensional free Dirac fermion, which has a $U(1)$ symmetry and time-reversal symmetry $\mathsf{T}$. Consider the partition function $\calZ[A]$ under the $U(1)$ gauge field $A$, then 
\be
\calZ[A]=|\calZ[A]|\exp(\im \eta[A]/2). 
\ee
Here, $\eta[A]$ is the eta invariant~\cite{Witten:2015aba}, which is roughly the $U(1)$ level-$1$ Chern--Simons action but is gauge-invariant modulo $4\pi$, and $U(1)$ itself has no 't Hooft anomaly. 
With the background $U(1)$ gauge field, the time-reversal symmetry is broken because 
\be
\calZ[\mathsf{T}\cdot A]=\calZ[A]\exp\left(-{\im\over 4\pi}\int A\diff A\right). 
\ee
That is, $U(1)$ and $\mathsf{T}$ has a mixed 't Hooft anomaly and it is characterized by the Chern-Simons action. Now, we consider the three-dimensional manifold of the form $M^3=M^2\times S^1$, and let $S^1$ be small enough. 
We want to check the fate of above anomaly in the two-dimensional effective theory, whose symmetry is $U(1)$ and $\mathsf{T}$. To gauge the $U(1)$ symmetry of this two-dimensional theory, we set $A=A_1\diff x^1+A_2\diff x^2$ with $x^3$-independent $A_i$, which is a $U(1)$ connection on $M^2$. The anomaly vanishes with this $U(1)$ gauge field $A$, since $A\diff A=0$. As we can see in this example, the anomaly is characterized by a topological invariant of the background gauge field $A$, and if we make $A$ be independent of one compactified direction the topological invariant vanishes identically. 
Physical interpretation is that thermal fluctuation appears after circle compactification and information of topology is lost because of it. 
This observation seems to be generic and shows the fundamental difficulty in making a connection between vacuum structures of the original theory and the circle-compactified theory from the viewpoint of 't Hooft anomaly matching. 

What is recently found is that if the 't Hooft anomaly involves a one-form symmetry then it does survive even at finite temperatures~\cite{Gaiotto:2017yup}. The known example is the $SU(N)$ Yang--Mills theory at $\theta=\pi$. 
$SU(N)$ Yang--Mills theory has the $\mathbb{Z}_N$ one-form symmetry that acts on Wilon lines. At $\theta=\pi$ the theory also has the time-reversal symmetry $\mathsf{T}$. 
If we consider the partition function $\calZ_{\theta=\pi}[B]$ with the background $\mathbb{Z}_N$ two-form gauge field $B$ for the center symmetry, the time-reversal symmetry is broken:
\be
\calZ_{\theta=\pi}[\mathsf{T}\cdot B]=\calZ_{\theta=\pi}[B]\exp\left({\im N\over 4\pi}\int B\wedge B\right). 
\ee
Therefore, there is a mixed 't Hooft anomaly between the $\mathbb{Z}_N$ one-form symmetry and time-reversal symmetry. At $\theta=\pi$, either of them must be broken spontaneously if we assume the mass gap at $\theta=\pi$.

Now, we compactify one direction and set the four-dimensional manifold as $M^4=M^3\times S^1$, where the size $L$ of $S^1$ is sufficiently small, i.e., the temperature $L^{-1}$ is sufficiently high.  
In this case, in addition to $\mathbb{Z}_N$ one-form symmetry, there exists $\mathbb{Z}_N$ zero-form symmetry that acts on Polyakov loop $\Phi=\mathrm{tr}[\mathcal{P}\exp\im\oint_{S^1}a]$. 
In order to gauge these $\mathbb{Z}_N$ symmetries, we introduce the $\mathbb{Z}_N$ two-form gauge field $B^{(2)}$ and also the $\mathbb{Z}_N$ one-form gauge field $B^{(1)}$. 
In the four-dimensional language, these gauge fields for three-dimensional effective theory can be regarded as 
\be
B=B^{(2)}+B^{(1)}\wedge L^{-1}\diff x^4. 
\ee
Substituting this form into the anomaly relation, we obtain 
\be
\calZ_{\theta=\pi}[\mathsf{T}\cdot B]=\calZ_{\theta=\pi}[B]\exp\left({\im N\over 2\pi}\int B^{(2)}\wedge B^{(1)}\right). 
\ee
This suggests that there is a mixed 't Hooft anomaly among $\mathbb{Z}_N$ zero-form, $\mathbb{Z}_N$ one-form, and time-reversal symmetries. 
Even at finite temperatures, one of these three symmetries must be spontaneously broken. 
The intuitive difference between anomalies involving only ordinary symmetry and containing one-form symmetries is the following: If the anomaly involves one-form symmetry, the line operator wrapping around $S^1$ is affected by the compactified direction even if 
that direction is small, and information of topology survives in the circle-compactified theory. 
This provides positive support to the idea of adiabatic continuity for Yang-Mills theory with adjoint matters~\cite{Unsal:2007vu, Kovtun:2007py, Unsal:2007jx, Unsal:2008ch, Shifman:2008ja, Shifman:2009tp} because we can claim that vacuum structures of the original and circle-compactified theories are controlled by the same 't Hooft anomaly. 

What happens if the gauge theory contains some matter fields not in the adjoint representation like quantum chromodynamics (QCD)? Typically, such theories do not have one-form symmetries, but still its vacuum property is sometimes constrained by 't Hooft anomalies. 
The same situation occurs in some two-dimensional nonlinear sigma models, such as the $\mathbb{C}P^{N-1}$ model, which has no one-form symmetries and an 't Hooft anomaly exists. 
There is a circumstantial evidence that 't Hooft anomalies should survive under circle compactifications even for these cases from the viewpoint of adiabatic continuity: Adiabatic continuity of two-dimensional sigma models seems to be valid under specific boundary conditions on $\mathbb{R}\times S^1$ \cite{Dunne:2012ae, Dunne:2012zk,  Cherman:2013yfa, Misumi:2014jua, Misumi:2014bsa}, and it is rigorously proven for $\mathbb{C}P^{N-1}$ or $O(N)$ sigma models in large-$N$ limit \cite{Sulejmanpasic:2016llc}. 
It is elucidated in the large-$N$ limit that the twisted boundary condition eliminates most of contributions to the partition function from high-energy states~\cite{Basar:2013sza,Sulejmanpasic:2016llc} as seen in Witten index of supersymmetric theories~\cite{Witten:1982im}, and the property of the vacuum is correctly captured at any size of circle compactification. 

In this paper, we positively answer the question whether there is any example in which an 't Hooft anomaly only of ordinary symmetries survives after circle compactification. 
We develop a systematic procedure generating the anomaly in the circle compactification starting from the anomaly of the original uncompactified theory, where the twisted boundary condition plays a pivotal role in our construction of anomaly.
Although no one-form symmetry exists, the above calculation for the case with one-form symmetry~\cite{Gaiotto:2017yup} gives us a strong motivation of our construction. 
Theories covered by our procedure contain $\mathbb{C}P^{N-1}$ sigma model with $\mathbb{Z}_N$ twisted boundary condition, massless QCD with $N_c=N_f$ with twisted boundary condition ($\mathbb{Z}_N$-QCD)~\cite{Kouno:2012zz, Sakai:2012ika, Kouno:2013zr, Kouno:2013mma, Poppitz:2013zqa, Iritani:2015ara}, and so on. 
Our systematic procedure is valid even away from large-$N$ limit so long as an 't Hooft anomaly exists. 

The paper is organized as follows. In Section~\ref{sec:formalism}, we establish a systematic procedure to compute the 't Hooft anomaly of  the circle compactified theory when the original one has no one-form symmetry. 
We there give a concrete construction of the anomaly, and the importance of appropriately twisted boundary condition is clarified. 
In Section~\ref{sec:CPN-1model}, we demonstrate our method in two-dimensional $\mathbb{C}P^{N-1}$ model. Starting from two-dimensional 't Hooft anomaly of $\mathbb{C}P^{N-1}$ model at $\theta=\pi$, we derive the anomaly of $\mathbb{Z}_N$-twisted $\mathbb{C}P^{N-1}$ model on $\mathbb{R}\times S^1$. 
In Section~\ref{sec:massless_qcd}, we discuss an anomaly of massless $\mathbb{Z}_N$-QCD and derive it starting from four-dimensional massless QCD. We compare the 't Hooft anomaly with results of previous studies, and discuss the application of anomaly matching to the phase diagram. 
Section~\ref{sec:conclusion} is devoted to conclusion and discussion.

\section{Formalism}\label{sec:formalism}

In this section, we develop a systematic procedure for deriving an 't Hooft anomaly of circle-compactified theories starting from the 't Hooft anomaly of an original theory. 
Only when there exists one-form symmetry, it has been already well-understood that 't Hooft anomaly survives even at finite temperatures~\cite{Gaiotto:2017yup}. 
Our procedure given below derives the anomaly of circle-compactified theories when the original theory has no one-form symmetries. 
It turns out that the boundary condition twisted by global symmetry plays an important role for nonvanishing anomaly. 

\subsection{Systematic procedure for anomaly with circle compactification}\label{sec:systematic_procedure}

We consider a $(D+1)$-dimensional quantum field theory (QFT), and assume that the QFT has two symmetries $G$ and $H$ acting on its physical Hilbert space faithfully. 
We call $G$ as a flavor symmetry, and consider the case when $G=\widetilde{G}/\Gamma$ with $\widetilde{G}=SU(N)$ and $\Gamma=\mathbb{Z}_{N}$. It is straightforward to extend the discussion for the case $\widetilde{G}=SU(N_1)\times SU(N_2)\times \cdots \times U(1)^r$ and $\Gamma=\mathbb{Z}_{n_1}\times \mathbb{Z}_{n_2}\times \cdots \subset Z(\widetilde{G})$ (center of $\widetilde{G}$). However, it makes notations for the following discussion complicated, and we try to make our explanation as simple as possible.

We further assume that $G$ and $H$ have a mixed 't Hooft anomaly but $\widetilde{G}$ and $H$ have no anomaly. 
In order to see the anomaly, we introduce the background gauge field for the flavor symmetry $G$. $G$-gauge field consists of two ingredients~\cite{Kapustin:2014gua, Gaiotto:2014kfa, Aharony:2013hda}:
\begin{itemize}
\item $\widetilde{G}$-gauge field $A$ that is locally a one-form in $(D+1)$ dimensions.
\item $\Gamma$-gauge field $B$ that is locally a two-form in $(D+1)$ dimensions. 
\end{itemize}
The physical interpretation is that the electric $\Gamma$ one-form symmetry appears after gauging $\widetilde{G}$ and its charged object is the Wilson line 
\be
W(\mathcal{C})=\mathrm{tr}\left[\mathcal{P}\exp\left(\im \oint_{\mathcal{C}} A\right)\right]. 
\ee
Gauging this one-form symmetry by $B$, the Wilson line is no longer a genuine line operator and we obtain $G$-gauge theory instead of $\widetilde{G}$-gauge theory. 
We denote the partition function of this QFT under the background $G$-gauge field as $\calZ[(A,B)]$. The above assumption on anomaly implies that 
\be
\calZ[h\cdot (A,B)]=\calZ[(A,B)] \exp\left(\im \mathcal{A}_h[B]\right), 
\label{eq:anomaly_D+1}
\ee
where $h\in H$, $h\cdot (A,B)$ is the $H$-transformation of $G$-gauge field $(A,B)$, and $\mathcal{A}_h[B]$ is a $(D+1)$-dimensional topological $\Gamma$-gauge theory\footnote{Generally speaking, the anomaly $\mathcal{A}_h$ can be a local $G$-gauge invariant functional of $A$ and $B$, $\mathcal{A}_h[A,B]$, such that $\mathcal{A}_h[A,0]\equiv 0$. For simplicity, we further assume the anomaly $\mathcal{A}_h$ depends only on $B$, but such extensions are inevitable if we consider the case when the dimension $D+1$ is odd. } determined by $h\in H$. 
For some $h\in H$, the anomaly $\mathcal{A}_h[B]$ cannot be canceled by variations of local counterterms. 
Under this setup, we will derive the 't Hooft anomaly of $D$-dimensional effective theory when one of the direction is compactified to a small circle.

In order to consider the circle compactification, we set $(D+1)$-dimensional manifolds as $M^{D+1}=M^D\times S^1$, and the size of its $D$-dimensional part $M^D$ is much larger than the circle $S^1$ of size $L$. At this stage, all the fields of QFT obey the periodic boundary condition along $S^1$. 
To describe the theory at first, we turn off $B$ and $D$-dimensional components of $A$, i.e., $A=A_{D+1}\diff x^{D+1}$. 
In this process, we can still fix the Polyakov-loop matrix of $A=: A_{\mathrm{cl}}$ along $S^1$ to a nontrivial one, and denote it as 
\be
\Omega=\mathcal{P}\exp\left(\im \int_0^L A_{\mathrm{cl}}\right). 
\ee
For each $\Omega\in\widetilde{G}$ that is uniform on $M^D$, we obtain a $D$-dimensional QFT on $M^D$, and denote its partition function as $\calZ_{\Omega}$. This is equivalent to imposing a twisted boundary condition on fields of QFT along $S^1$ by performing a boundary-condition-changing $\widetilde{G}$-gauge transformation, but we keep the periodic boundary condition with nontrivial holonomy $\Omega$ during our explanation of the general strategy. 
Before introducing the two-form gauge field $B$, we originally have $(D+1)$-dimensional one-form symmetry $\Gamma$, and it induces $D$-dimensional zero-form and one-form symmetries $\Gamma$ after circle compactification when $A$ is dynamical. 
The zero-form symmetry $\Gamma(=\mathbb{Z}_N)$ acts as $\mathrm{tr}(\Omega^n)\mapsto \omega^n \mathrm{tr}(\Omega^n)$ with some $\omega\in \Gamma\setminus \{\bm{1}\}$, and we thus identify\footnote{This identification is not gauge-invariant since $\Omega$ itself is not. To justify it, we regard that the Polyakov gauge is taken for $\widetilde{G}$-gauge field $A$. } its action on $\Omega$ itself as $\Omega\mapsto \omega \Omega$. 
However, since we define the theory $\calZ_{\Omega}$ by fixing the $\widetilde{G}$ holonomy $\Omega$, the above transformation $\Omega\mapsto \omega \Omega$ maps one theory $\calZ_{\Omega}$ to  another theory $\calZ_{\omega\Omega}$: It is not the symmetry of $\calZ_{\Omega}$.

In order to have a nontrivial anomaly on $M^D$, we need to have a symmetry involving the above zero-form transformation, $\Omega\mapsto \omega\Omega$. We specify the $\widetilde{G}$ holonomy $\Omega$ such that there exists $S\in \widetilde{G}$ satisfying 
\be
S \Omega S^{-1}=\omega \Omega\,.
\label{eq:shift_symmetry}
\ee 
Since $S$ is not an element of the center $Z(\widetilde{G})$ of $\widetilde{G}$ by definition, $S\not \in \Gamma$. Recall that $G=\widetilde{G}/\Gamma$ acts faithfully on the physical Hilbert space, and it means that $S$ generates a faithful symmetry of QFT, which we call a ``shift symmetry''. 
When the shift symmetry $S$ acts on fields, the holonomy matrix $\Omega$ is changed to $S\Omega S^{-1}$. 
The requirement (\ref{eq:shift_symmetry}) states that the symmetry generated by $S$ is intertwined with the zero-form symmetry $\Gamma$, $\Omega\mapsto \omega^{-1}\Omega$, in order to maintain the holonomy $\Omega$, and the symmetry of $\calZ_{\Omega}$ is obtained: We denote this zero-form symmetry $\Gamma$ generated by $S$ as $\Gamma_S$ in order to distinguish it from the original one, $\Gamma\subset Z(\widetilde{G})$. 
Because of (\ref{eq:shift_symmetry}), $\Omega$ cannot be proportional to the identity matrix: Typical example of $\Omega$ and $S$ satisfying (\ref{eq:shift_symmetry}) is ($\omega=\mathrm{e}^{2\pi \im/N}$)
\be
\Omega= \omega^{-(N-1)/2}\left(\begin{array}{ccccc}
1&0&0&\cdots&0\\
0&\omega&0&\cdots&0\\
0&0&\omega^2&\cdots &0\\
\vdots&\vdots&\vdots& &\vdots\\
0&0&0&\cdots&\omega^{N-1}
\end{array}\right), \;\; 
S=\left(\begin{array}{ccccc}
0&1&0&\cdots&0\\
0&0&1&\cdots&0\\
 \vdots&\vdots &\vdots & &\vdots\\
0&0&0&\cdots&1\\
1&0&0&\cdots&0
\end{array}\right).
\label{eq:formalism_Omega_Shift}
\ee 
Since the flavor symmetry of $D$-dimensional theory must commute with $\Omega$, flavor symmetry $\widetilde{G}$ might be explicitly broken to a maximal Abelian subgroup $\widetilde{K}$ as $\widetilde{G}\to \widetilde{K}$. Symmetry with the faithful representation is again given by the quotient $K=\widetilde{K}/\Gamma$. 
Let us assume that $H$ is not explicitly broken by fixing $\Omega$, then the $D$-dimensional effective theory $\calZ_{\Omega}$ has three symmetries; shift symmetry $\Gamma_S$, flavor symmetry $K$, and $H$.  

Let us try to introduce the background gauge fields for $\Gamma_S$ and $K$. We denote the $\Gamma_S$-gauge field as $B^{(1)}$ that is locally a one-form on $M^D$.  
The $K$-gauge field consists of two ingredients: 
\begin{itemize}
\item $\widetilde{K}$-gauge field $A_K$ that is locally a one-form on $M^D$.
\item $\Gamma$-gauge field $B^{(2)}$ that is locally a two-form on $M^D$. 
\end{itemize}
When $\widetilde{K}$ is gauged, the $D$-dimensional one-form symmetry $\Gamma$ emerges, so we can introduce $B^{(2)}$ which is a two-form on $M^D$. 
Using these gauge fields, we define the $G$-gauge field on $M^D\times S^1$ as 
\be
A=A_K+B^{(1)}+A_{\mathrm{cl}},\;\; B=B^{(2)}+B^{(1)}\wedge {L^{-1}\diff x^{D+1}}. 
\label{eq:two-form_D-dim}
\ee
The field $B^{(1)}$ in the expression (\ref{eq:two-form_D-dim}) may require some explanations: When it appears in $A$, it is regarded as a gauge field for the subgroup $\Gamma_S\subset \widetilde{G}$ of the flavor symmetry, while it should be regarded as a gauge field for the zero-form symmetry $\Gamma$ induced from the one-form symmetry when it appears inside $B$. 
We use this slightly abused notation in order to emphasize that these two transformations are intertwined as a symmetry of $\calZ_{\Omega}$. 
Let $\calZ_{\Omega}[(A_K,B^{(1)},B^{(2)})]$ be the partition function with the background gauge fields for $\Gamma_S$ and $K$, then the given construction of the $D$-dimensional theory shows that  
\be
\calZ_{\Omega}[(A_K,B^{(1)},B^{(2)})]=\calZ[(A,B)], 
\label{eq:relation_D_D+1}
\ee
where $A$ and $B$ are given in (\ref{eq:two-form_D-dim}). 

We can now derive the mixed $D$-dimensional anomaly among $\Gamma_{S}$, $K$, and $H$ as follows: Using the $(D+1)$-dimensional expression of the $D$-dimensional theory, (\ref{eq:two-form_D-dim}) and (\ref{eq:relation_D_D+1}), we obtain 
\bea
&&\calZ_{\Omega}[h\cdot (A_K,B^{(1)},B^{(2)})]\nonumber\\
&=&\calZ[h\cdot(A,B)]\nonumber\\
&=&\calZ[(A,B)] \exp\left(\im\mathcal{A}_h[B]\right) \nonumber\\
&=&\calZ_{\Omega}[(A_K,B^{(1)},B^{(2)})] \exp\left(\im\mathcal{A}_h[B^{(2)}+B^{(1)}\wedge L^{-1} \diff x^{D+1}]\right). 
\label{eq:anomaly_D-dim}
\eea
Here, we use the anomaly relation (\ref{eq:anomaly_D+1}) in $(D+1)$ dimensions. 
Since $B^{(2)}$ and $B^{(1)}$ have no dependence on $x^{D+1}$, the compactified direction $S^1$ of $\mathcal{A}_h[B^{(2)}+B^{(1)}\wedge L^{-1}\diff x^{D+1}]$ can be integrated out in a trivial manner. As a result, $\mathcal{A}_h$ becomes the $D$-dimensional topological $\Gamma$-gauge theory of $B^{(2)}$ and $B^{(1)}$, and this defines the anomaly of the $D$-dimensional QFT $\calZ_{\Omega}$. 
That is, the $D$-dimensional QFT $\calZ_{\Omega}$ has a mixed 't Hooft anomaly among $\Gamma_S$, $K$, and $H$, and the trivial gapped state is forbidden. 

\subsection{Comments on choice of the background holonomy}
\label{sec:comments_holonomy}

One of the most important part in our construction of $D$-dimensional anomaly via circle compactification is the choice of nontrivial holonomy background $\Omega$. 
In order to clarify the importance of the condition (\ref{eq:shift_symmetry}), let us take the trivial one $\Omega=\bm{1}$ as a ``bad'' example. 

Whether or not (\ref{eq:shift_symmetry}) is satisfied, the zero-form transformation $\Gamma$ acts on holonomy as $\Omega\mapsto \omega \Omega$. However, any elements $g\in \widetilde{G}$ commutes with $\Omega$, and then this zero-form transformation on the Polyakov loop has no connection with the symmetry of $D$-dimensional system. 
Since it is not a faithful symmetry of the system, we must set $B^{(1)}=0$ in the trivial choice $\Omega=\bm{1}$. Instead, the flavor symmetry $\widetilde{G}$ is unbroken, and the corresponding gauge field on $M^D$ is denoted as $A_G$. 
As a result, the anomaly relation (\ref{eq:anomaly_D-dim}) becomes 
\be
\calZ_{\bm{1}}[h\cdot(A_G, 0, B^{(2)})]=\calZ_{\bm{1}}[(A_G,0,B^{(2)})]\exp\left(\im \mathcal{A}_h[B^{(2)}]\right). 
\ee
Since $B^{(2)}$ is a gauge-field on $M^D$ while $\mathcal{A}_h$ is a $(D+1)$-dimensional topological theory, we obtain $\mathcal{A}_h[B^{(2)}]=0$. 
This means that the trivial boundary condition eliminates the anomaly of $(D+1)$ dimensions. Two-form gauge fields $B^{(2)}$ on $M^D$ are not enough for anomaly. 

In our construction, the zero-form transformation $\Gamma$ on Polyakov-loop $\Omega$ is translated into the faithful symmetry generated by $S$ on fields of QFT via (\ref{eq:shift_symmetry}). The appearance of the faithful zero-form symmetry $\Gamma_S$ allows us to introduce the $\Gamma_S$-gauge field $B^{(1)}$. 
Since $\Gamma_S$ and $\Gamma$ is intertwined in $(D+1)$ dimensions, it is built into the $(D+1)$-dimensional two-form gauge field $B$ as a form $B^{(1)}\wedge L^{-1}\diff x^{D+1}$. Therefore, the 't Hooft anomaly can survive even if the original theory has no one-form symmetry.

\section{$\mathbb{Z}_N$-twisted $\mathbb{C}P^{N-1}$ sigma model}\label{sec:CPN-1model}

As a demonstration of the systematic procedure in Sec.~\ref{sec:formalism}, we calculate the anomaly of $\mathbb{Z}_N$-twisted $\mathbb{C}P^{N-1}$ model starting from the two-dimensional 't Hooft anomaly.  
Two-dimensional $\mathbb{C}P^{N-1}$ sigma model can be realized as a gauged linear sigma model,  
\be
S=\int \diff^2 x \left[{1\over 2}|(\p_{\mu}+\im a_{\mu})\vec{z}|^2+{\lambda\over 4}(|\vec{z}|^2-\mu^2)^2\right]-{\im \theta\over 2\pi}\int \diff a, 
\label{eq:2d_CPN-1}
\ee
where $\vec{z}=(z_1,\ldots,z_N)$ is an $N$-component complex vector-valued fields, and $a$ is the $U(1)$-gauge field\footnote{In this paper, we use the lower-case for dynamical gauge fields and the upper-case for background, or classical, gauge fields. }. 
We study theta-dependence of this theory from the viewpoint of anomaly. We start with the two-dimensional discussion~\cite{Komargodski:2017dmc} first, and move to the circle compactification with $\mathbb{Z}_N$-twisted boundary condition. 
In order to understand the formalism better, we follow each of the steps explained in Sec.~\ref{sec:systematic_procedure} in detail. 

\subsection{'t Hooft anomaly and global inconsistency in two dimensions} 

The symmetry of this theory (\ref{eq:2d_CPN-1}) consists of the following~\cite{Komargodski:2017dmc}: 
\begin{itemize}
\item Flavor symmetry, $SU(N)/\mathbb{Z}_N$, which is given by $\vec{z}\mapsto U\vec{z}$ with $U\in SU(N)$. 
\item Time-reversal symmetry $\mathsf{T}$ at $\theta=0,\pi$. 
\end{itemize}
Since the $U(1)$ symmetry is gauged, the center elements of $SU(N)$ cannot act faithfully on gauge-invariant operators. The flavor symmetry with the faithful representation is thus given by $SU(N)/\mathbb{Z}_N$. 
In the notation used in Sec.~\ref{sec:formalism}, we have the following correspondence: $G=SU(N)/\mathbb{Z}_N$, $\widetilde{G}=SU(N)$, $\Gamma=\mathbb{Z}_N$, and $H=\mathsf{T}$. 

We introduce the background gauge field for flavor $SU(N)/\mathbb{Z}_N$ symmetry. Such a background gauge field consists of two ingredients: $SU(N)$ gauge field $A$ and $\mathbb{Z}_N$ two-form gauge field $B$. To explain it, let us first gauge the $SU(N)$ symmetry, then the action obtained by the minimal coupling becomes 
\be
S_{\mathrm{gauged}}=\int \diff^2 x \left[{1\over 2}|(\p_{\mu}+\im a_{\mu}-\im A_{\mu})\vec{z}|^2+V(|\vec{z}|^2)\right]-{\im \theta\over 2\pi}\int \diff a, 
\ee
where $V(|\vec{z}|^2)={\lambda\over 4}(|\vec{z}|^2-\mu^2)^2$. 
Flavor symmetry $SU(N)$ is gauged and no longer a global symmetry, but the theory acquires the one-form symmetry: Considering the $U(1)$ and $SU(N)$ Wilson lines, 
\be
W_{U(1)}(\mathcal{C})=\exp\left(\im \int_{\mathcal{C}} a\right),\; W_{SU(N)}(\mathcal{C})=\mathrm{tr}\left[\mathcal{P}\exp\left(\im \int_{\mathcal{C}} A\right)\right], 
\ee
then the theory has a symmetry under the simultaneous $\mathbb{Z}_N$ rotation, 
\be
W_{U(1)}(\mathcal{C})\mapsto \mathrm{e}^{2\pi \im/N}W_{U(1)}(\mathcal{C}),\; W_{SU(N)}(\mathcal{C})\mapsto \mathrm{e}^{2\pi \im/N}W_{SU(N)}(\mathcal{C}). 
\ee
The Wilson lines charged under this $\mathbb{Z}_N$ one-form symmetry must be dropped from the spectrum of genuine line operators if we appropriately gauge the flavor $SU(N)/\mathbb{Z}_N$ symmetry~\cite{Kapustin:2014gua, Gaiotto:2014kfa}. For this purpose, we introduce the $\mathbb{Z}_N$ two-form gauge field $B$, and then we obtain 
\be
\calZ_{\theta}[(A,B)]=\int \Diff a\Diff\vec{z} \exp\left[-\int \diff^2 x \left({1\over 2}|(\p_{\mu}+\im a_{\mu}-\im A_{\mu})\vec{z}|^2+V(|\vec{z}|^2)\right)+{\im \theta\over 2\pi}\int (\diff a+B)\right]. 
\ee

At $\theta=\pi$, we consider the time-reversal transformation under the background flavor gauge field $(A,B)$, and we obtain~\cite{Komargodski:2017dmc}
\be
\calZ_{\pi}[\mathsf{T}\cdot (A,B)]=\calZ_{\pi}[(A,B)]\mathrm{e}^{-\im \int B}. 
\label{eq:CPN-1_anomaly_2d}
\ee 
We should check whether this anomaly is genuine or fake, so we consider whether it can be canceled by local counter terms of $B$. 
The topological $\mathbb{Z}_N$ two-form gauge theory is given by $\im k\int B$ with some integer $k$ modulo $N$, and it is a candidate for the counterterm. 
The $\mathsf{T}$ transformation after adding this counterterm behaves as 
\be
\calZ_{\pi}[\mathsf{T}\cdot (A,B)]\exp\left(-\im k\int \mathsf{T}\cdot B\right)=\calZ_{\pi}[(A,B)]\exp\left(-\im k\int B\right)\mathrm{e}^{\im (2k-1)\int B}. 
\ee
Thus, anomaly is fake if and only if $2k=1$ modulo $N$. For even $N$, this is impossible and we find the 't Hooft anomaly between the flavor and time-reversal symmetries. 

For odd $N$, we can eliminate the anomaly by choosing $k=(N+1)/2$ modulo $N$, and no 't Hooft anomaly exists. If we do the same computation at $\theta=0$, the time-reversal symmetry is respected by choosing $k=0$, and thus there is no common counterterm $k$ that respects $\mathsf{T}$ both at $\theta=0,\pi$. 
This is the global inconsistency condition, and we can derive a nontrivial consequence although it is slightly weaker than 't Hooft anomaly matching (see Ref.~\cite{Kikuchi:2017pcp} for global inconsistency condition). 

In two dimensions, Coleman--Mermin--Wagner theorem~\cite{Coleman:1973ci, mermin1966absence} tells us that flavor symmetry with a continuous parameter cannot be broken. 
This naturally gives us the nonperturbative data that time-reversal symmetry at $\theta=\pi$ is spontaneously broken for two-dimensional $\mathbb{C}P^{N-1}$ model. 

\subsection{$\mathbb{Z}_N$-twisted $\mathbb{C}P^{N-1}$ model and its anomaly}

We consider the circle compactification from $\mathbb{R}^2$ to $\mathbb{R}\times S^1\ni (x^1,x^2)$, where the circumference $L$ of the circle $S^1$, i.e. $x^2\sim x^2+L$,  is regarded to be small.  We impose the $\mathbb{Z}_N$-twisted boundary condition, 
\be
\vec{z}(x^1,x^2+L)=\Omega \vec{z}(x^1,x^2), 
\label{eq:twisted_CP_boundary}
\ee
where ($\omega=\mathrm{e}^{2\pi \im/N}$)
\be
\Omega=\mathrm{diag}(1,\omega, \ldots, \omega^{N-1}). 
\ee
We call this as $\mathbb{Z}_N$-twisted $\mathbb{C}P^{N-1}$ sigma model, and we denote its partition function at $\theta$ as $\calZ_{\theta,\Omega}$. 

For our purpose, it is better to regard this twisting matrix $\Omega$ as a holonomy of $SU(N)$ gauge field $A$ along compactified direction. Indeed, the boundary-condition-changing $SU(N)$ gauge transformation makes the boundary condition of $\vec{z}$ periodic (up to $U(1)$ gauge symmetry) and the price to be paid is the background $SU(N)$ holonomy $\propto\Omega$. 
When $SU(N)$ is gauged, there is a $\mathbb{Z}_N$ one-form symmetry and it induces the $\mathbb{Z}_N$ zero-form symmetry, $\Omega\mapsto \omega \Omega$. 
What is special for this choice of the nontrivial holonomy $\Omega$ is that the above $\mathbb{Z}_N$ transformation induces the symmetry given by 
\be
\vec{z}=\left(\begin{array}{c}
z_1\\ z_2\\ \vdots\\ z_{N-1}\\ z_{N}
\end{array}\right)
\mapsto S\vec{z}=\left(\begin{array}{c}
z_2\\ z_3\\ \vdots\\ z_{N}\\ z_{1}
\end{array}\right), 
\;\; S=\left(\begin{array}{ccccc}
0&1&0&\cdots&0\\
0&0&1&\cdots&0\\
 \vdots&\vdots &\vdots & &\vdots\\
0&0&0&\cdots&1\\
1&0&0&\cdots&0
\end{array}\right). 
\ee
We call this a shift symmetry, and it is a faithful transformation on the physical spectrum. As an example, a gauge-invariant operator $|z_1|^2$ is mapped to another gauge-invariant operator $|z_2|^2$. 
When the transformation $S$ is performed as $\vec{z}\mapsto \vec{z'}=S\vec{z}$, the boundary condition for the transformed field becomes
\be
\vec{z'}(x^1,x^2+L)=S\Omega S^{-1} \vec{z'}(x^1,x^2)=\omega \Omega \vec{z'}(x^1,x^2). 
\ee 
That is, the $\mathbb{Z}_N$ zero-form symmetry on the Polyakov loop $\Omega$ is intertwined with the shift symmetry $\mathbb{Z}_N$ generated by $S$ in order to maintain the boundary condition (\ref{eq:twisted_CP_boundary}), and we call it $(\mathbb{Z}_N)_S$. The symmetry $(\mathbb{Z}_N)_S$ acts on local operators on $\mathbb{R}$ as 
\be
\vec{z}\mapsto S\vec{z},\;\; \exp\left(\im \int_{S^1}a\right)\mapsto \omega^{-1}\exp\left(\im \int_{S^1}a\right). 
\ee
We give a short summary of the situation: The compactified theory obtained here has a $(\mathbb{Z}_N)_S$ zero-form symmetry, and it is induced by the $\mathbb{Z}_N$ one-form symmetry in two dimensions when $SU(N)$ is gauged. 
Continuous part of the flavor symmetry is explicitly broken to $U(1)^{N-1}/\mathbb{Z}_N$, but it is not relevant for the following discussion and we do not introduce gauge fields for it. 

As a result, the 't Hooft anomaly (or global inconsistency) in two dimensions has the same meaning in the $\mathbb{Z}_N$-twisted $\mathbb{C}P^{N-1}$ model on $\mathbb{R}\times S^1$. 
The $\mathbb{Z}_N$-twisted $\mathbb{C}P^{N-1}$ model has the shift symmetry $(\mathbb{Z}_N)_S$ and the time-reversal symmetry $\mathsf{T}$ at $\theta=0,\pi$. 
We introduce the $\mathbb{Z}_N$ one-form gauge field $B^{(1)}$ for gauging $(\mathbb{Z}_N)_S$, which is independent of $x^2$. 
Since $(\mathbb{Z}_N)_S$ is intertwined with the $\mathbb{Z}_N$ zero-form symmetry acting on the $SU(N)$ and $U(1)$ Polyakov loops, we can embed it into the $\mathbb{Z}_N$ two-form gauge field $B$ in the two-dimensional language by setting $B=B^{(1)}\wedge L^{-1}\diff x^2$. 
The two-dimensional anomaly (\ref{eq:CPN-1_anomaly_2d}) tells us that 
\bea
\calZ_{\pi,\Omega}[\mathsf{T}\cdot B^{(1)}]&=&\calZ_{\pi,\Omega}[B^{(1)}]\exp\left(-\im \int B^{(1)}\int_0^L L^{-1}\diff x^2\right)\nonumber\\
&=&\calZ_{\pi,\Omega}[B^{(1)}]\exp\left(-\im \int B^{(1)}\right). 
\label{eq:CPN-1_anomaly_1d}
\eea
We find that $(\mathbb{Z}_N)_S$ and $\mathsf{T}$ at $\theta=\pi$ has an 't Hooft anomaly (or global inconsistency depending on even or odd $N$), and either of them must be spontaneously broken. 

For usual periodic boundary condition, it means that we put $A=0$ and thus there is no room to introduce $B$. Therefore, we cannot obtain 't Hooft anomaly in such cases. The emergence of $\mathbb{Z}_N$ symmetry by the twisted boundary condition is essential for a deep connection with two-dimensional anomaly.

\subsection{Comparison with previous studies and Discussion}

The $\theta$-angle dependence of $\mathbb{C}P^{N-1}$ model in two dimensions is studied in large-$N$ limit~\cite{Witten:1978bc, Affleck:1979gy}, and the ground state energy should behave as 
\be
E(\theta)\propto \min_{k\in\mathbb{Z}}{1\over N}(\theta+2\pi k)^2. 
\label{eq:2d_CPN-1_theta}
\ee
This behavior matches the 't Hooft anomaly (\ref{eq:CPN-1_anomaly_2d}), because the time-reversal symmetry is spontaneously broken at $\theta=\pi$. 
Our derivation of the anomaly (\ref{eq:CPN-1_anomaly_1d}) for $\mathbb{Z}_N$-twisted $\mathbb{C}P^{N-1}$ model claims that the same multi-branch structure would naturally appear under adiabatic circle compactification. 

Indeed, $\theta$-dependence of $\mathbb{Z}_N$-twisted $\mathbb{C}P^{N-1}$ model is studied in Refs.~\cite{Dunne:2012ae, Dunne:2012zk}. Under the twisted boundary condition, there are $N$ types of fractional instantons which has the topological charge $1/N$. As a result, the quasi-ground states are composed of $N$ states and the $k$-th ground-state energy behaves as 
\be
E_k(\theta)\propto -N\cos\left({\theta+2\pi k\over N}\right). 
\label{eq:1d_CPN-1_theta}
\ee 
The ground state energy is thus given by minimum of these, 
\be
E(\theta)=\min_{k=1,\ldots,N}E_k(\theta). 
\ee
We can see that the time-reversal symmetry is spontaneously broken at $\theta=\pi$, which satisfies matching of 't Hooft anomaly or global inconsistency (\ref{eq:CPN-1_anomaly_1d}). 
What we have shown in this paper is that these two behaviors (\ref{eq:2d_CPN-1_theta}) and (\ref{eq:1d_CPN-1_theta}) are both consistent with anomalies, and those anomalies have essentially the same origin. 

We argue that this observation gives a positive support for the adiabatic continuity. For $\mathbb{Z}_N$-twisted $\mathbb{C}P^{N-1}$ model, it is rigorously shown that expectation values of any $SU(N)$ invariant operators does not depend on $L$ in the large-$N$ limit~\cite{Sulejmanpasic:2016llc}, and our consideration on anomaly gives a complementary and consistent analysis. 

\section{Massless $\mathbb{Z}_N$-QCD}\label{sec:massless_qcd}

As another demonstration, we consider a four-dimensional example: $SU(N)$ Yang-Mills theory with $N$ massless Dirac fermions in the fundamental representation, i.e. massless QCD with $N_f=N_c=N$. 
The action of this theory is given by 
\be
S={1\over 2g^2}\int \mathrm{tr}(G_{\mathrm{c}}\wedge *G_{\mathrm{c}})+\int \diff^4 x\, \mathrm{tr}\left\{\overline{\Psi}\gamma_{\mu}D_{\mu}(a)\Psi\right\}, 
\label{eq:massless_qcd}
\ee
where $a$ is the $SU(N)$ color gauge field, $D(a)=\diff+\im\, a$ is the covariant derivative, $G_{\mathrm{c}}=\diff a+\im\, a\wedge a$ is the $SU(N)$-gauge field strength, and $\Psi$ are $N\times N$ matrix-valued Dirac fermions.   
The $SU(N)$ color group acts on $\Psi$ from left, and the $SU(N)$ flavor group acts on $\Psi$ from right, i.e., $\Psi\mapsto U_{\mathrm{c}} \Psi U_{\mathrm{f}}^{\dagger}$ for $(U_{\mathrm{c}},U_{\mathrm{f}})\in SU(N)_{\mathrm{color}}\times SU(N)_{\mathrm{flavor}}$; the quark field $\Psi$ is in the bifundamental representation of the color and flavor groups. 

This theory possesses various symmetries, but we pay attention only to the vector-like flavor symmetry, $SU(N)_{\mathrm{flavor}}/(\mathbb{Z}_N)_{\mathrm{color}-\mathrm{flavor}}$, and the anomaly-free discrete subgroup of the axial symmetry, $(\mathbb{Z}_{2N})_{\mathrm{axial}}$, in our demonstration. 
The complete analysis involving other symmetries will be discussed at future opportunity. 
We first compute the 't Hooft anomaly of the above symmetries in four dimensions. Using the four-dimensional computation, we derive the anomaly of the circle-compactified theory with the $\mathbb{Z}_N$-twisted boundary condition, $\mathbb{Z}_N$-QCD. 

Our discussion can be generalized to the case when color and flavor have different numbers $N_c\not=N_f$ so long as they have a nontrivial common divisor, $\gcd(N_c,N_f)>1$. For notational simplicity, we only consider the case $N_c=N_f=N$ in this paper. 

\subsection{Four-dimensional 't Hooft anomaly of massless $N$-flavor QCD}\label{sec:qcd_4d}

Here, we start with explanation on the vector-like flavor symmetry $SU(N)_{\mathrm{flavor}}/(\mathbb{Z}_N)_{\mathrm{color-flavor}}$ and $(\mathbb{Z}_{2N})_{\mathrm{axial}}$. 
In the notation used in Sec.~\ref{sec:formalism}, we have the following correspondence: $G=SU(N)_{\mathrm{flavor}}/(\mathbb{Z}_N)_{\mathrm{color-flavor}}$, $\widetilde{G}=SU(N)_{\mathrm{flavor}}$, $\Gamma=(\mathbb{Z}_N)_{\mathrm{color-flavor}}$, and $H=(\mathbb{Z}_{2N})_{\mathrm{axial}}$. 

Quark field $\Psi$ is in the bifundamental representation, $\Psi\mapsto U_{\mathrm{c}} \Psi U_{\mathrm{f}}^{\dagger}$ for $(U_{\mathrm{c}},U_{\mathrm{f}})\in SU(N)_{\mathrm{color}}\times SU(N)_{\mathrm{flavor}}$, and thus the subgroup $(\mathbb{Z}_N)_{\mathrm{color-flavor}}$ generated by $(\omega \bm{1}_N,\omega \bm{1}_{N})\in SU(N)_{\mathrm{color}}\times SU(N)_{\mathrm{flavor}}$ does not act on $\Psi$ faithfully ($\omega=\mathrm{e}^{2\pi \im/N}$). 
Therefore, the group acting faithfully on $\Psi$ is $\left(SU(N)_{\mathrm{color}}\times SU(N)_{\mathrm{flavor}}\right)/(\mathbb{Z}_N)_{\mathrm{color-flavor}}$. 
Since the color group symmetry $SU(N)_{\mathrm{color}}$ is gauged, the flavor symmetry with the faithful representation on the physical Hilbert space is given by $SU(N)_{\mathrm{flavor}}/(\mathbb{Z}_N)_{\mathrm{color-flavor}}$. 

Since the quark field $\Psi$ is massless, there is a symmetry $U(1)_{\mathrm{axial}}$, $\Psi\mapsto \mathrm{e}^{\im \alpha\gamma_5}\Psi$, at the Lagrangian level, but the fermion integration measure $\Diff \overline{\Psi}\Diff \Psi$ generates the additional term $\im {2N\alpha\over 8\pi^2}\int \mathrm{tr}(G_{\mathrm{c}}\wedge G_{\mathrm{c}})$ due to the index theorem. Therefore, it is a symmetry only when $\alpha$ is quantized to $2\pi/2N$, and $U(1)_{\mathrm{axial}}$ is explicitly broken to $(\mathbb{Z}_{2N})_{\mathrm{axial}}$ by quantum anomaly.

We shall derive the mixed 't Hooft anomaly between $SU(N)_{\mathrm{flavor}}/(\mathbb{Z}_N)_{\mathrm{color-flavor}}$ and $(\mathbb{Z}_{2N})_{\mathrm{axial}}$, and we introduce the background gauge field of $SU(N)_{\mathrm{flavor}}/(\mathbb{Z}_N)_{\mathrm{color-flavor}}$ for that purpose. 
We first introduce the flavor $SU(N)$ gauge field $A$, then the minimal-coupling procedure\footnote{We can add arbitrary gauge-invariant terms made of $A$ to it, but it does not change the result in our case. We therefore neglect them for simplicity, but they can be important in other cases~\cite{Gaiotto:2017tne}. } changes the action (\ref{eq:massless_qcd}) as 
\be
S_{\mathrm{gauged}}={1\over 2g^2}\int \mathrm{tr}(G_{\mathrm{c}}\wedge *G_{\mathrm{c}})+\int \diff^4 x\, \mathrm{tr}\left\{\overline{\Psi}\gamma_{\mu}D_{\mu}(a,A)\Psi\right\}, 
\label{eq:massless_qcd_flavor}
\ee
where the covariant derivative is replaced by
\be
D_{\mu}(a,A)\Psi=\p_{\mu}\Psi+\im a_{\mu}\Psi-\im \Psi A_{\mu}. 
\ee
The theory (\ref{eq:massless_qcd_flavor}) has $\mathbb{Z}_N$ one-form symmetry that does not exist in the original massless QCD. 
It acts on the color and flavor Wilson lines, $W(\mathcal{C})_{\mathrm{color}}=\mathrm{tr}\left[\mathcal{P}\exp\im\oint_{\mathcal{C}} a\right]$ and $W(\mathcal{C})_{\mathrm{flavor}}=\mathrm{tr}\left[\mathcal{P}\exp\im\oint_{\mathcal{C}} A\right]$, as the simultaneous rotation of $\mathbb{Z}_N$ phase,
\be
W(\mathcal{C})_{\mathrm{color}}\mapsto \omega W(\mathcal{C})_{\mathrm{color}},\;\; 
W(\mathcal{C})_{\mathrm{flavor}}\mapsto \omega W(\mathcal{C})_{\mathrm{flavor}}. 
\label{eq:Zn_one-form_QCDbifund}
\ee
This symmetry does not arise in the $SU(N)_{\mathrm{flavor}}/(\mathbb{Z}_N)_{\mathrm{color-flavor}}$ gauge theories, and we have to introduce $\mathbb{Z}_N$ two-form gauge field $B$~\cite{Kapustin:2014gua, Gaiotto:2014kfa}, which is a $U(1)$ two-form gauge field satisfying
\be
N B+\diff C=0, 
\ee
with a certain $U(1)$ one-form gauge field $C$. This constraint respects the $U(1)$ one-form gauge invariance under $B\mapsto B+\diff \lambda$ and $C\mapsto C-N \lambda$, and respecting this gauge invariance prevents us from adding extra degrees of freedom to the theory.  We introduce the $U(N)$ gauge fields made of $SU(N)$ gauge fields $a$, $A$, and a $U(1)$ gauge field $C$, as 
\be
\widetilde{a}=a+{1\over N}C,\;\; \widetilde{A}=A+{1\over N}C, 
\ee
and define their gauge field strengths as 
\be
\calG_\mathrm{c}=\diff \widetilde{a}+\im\, \widetilde{a}\wedge \widetilde{a},\;\; 
\calG_\mathrm{f}=\diff \widetilde{A}+\im \widetilde{A}\wedge \widetilde{A}. 
\ee
These field strengths transform under the $U(1)$ one-form gauge transformation as $\mathcal{G}_\mathrm{c}\mapsto \mathcal{G}_\mathrm{c}-\diff \lambda$ and $\mathcal{G}_\mathrm{f}\mapsto \mathcal{G}_\mathrm{f}-\diff\lambda$, and we obtain the gauge invariant combinations, $\mathcal{G}_\mathrm{c}+B$ and $\mathcal{G}_\mathrm{f}+B$. This tells us that the introduction of $\mathbb{Z}_N$ two-form gauge field $B$ changes the action (\ref{eq:massless_qcd_flavor}) as 
\be
S_{\mathrm{gauged}}={1\over 2g^2}\int \mathrm{tr}\Bigl\{(\calG_{\mathrm{c}}+B)\wedge *(\calG_{\mathrm{c}}+B)\Bigr\}+\int \diff^4 x\, \mathrm{tr}\left\{\overline{\Psi}\gamma_{\mu}D_{\mu}(\widetilde{a},\widetilde{A})\Psi\right\}.
\label{eq:massless_qcd_flavor_ZN}
\ee
We can shortly summarize the set of above procedures as follows: Introducing the flavor $SU(N)$ gauge field $A$, the theory becomes $SU(N)\times SU(N)$ bifundamental QCD, which has $\mathbb{Z}_N$ one-form symmetry, so we again introduce the $\mathbb{Z}_N$ two-form gauge field $B$~\cite{Tanizaki:2017bam, Shimizu:2017asf}.

Let us perform the $\mathbb{Z}_{2N}$ axial rotation, then the Lagrangian is invariant again, but the action acquires the additional topological term due to the fermion measure: 
\be
\Delta S={\im\over 4\pi}\int \mathrm{tr}\Bigl\{(\mathcal{G}_\mathrm{c}+B)\wedge (\mathcal{G}_\mathrm{c}+B)\Bigr\}
+{\im\over 4\pi}\int \mathrm{tr}\Bigl\{(\mathcal{G}_\mathrm{f}+B)\wedge (\mathcal{G}_\mathrm{f}+B)\Bigr\}
=-{\im 2N\over 4\pi}\int B\wedge B. 
\ee
The last equality holds modulo $2\pi\im$\footnote{For details of this computation, the related calculations are available in Ref.~\cite{Tanizaki:2017bam} in the almost same convention.}. 
This means that the $(\mathbb{Z}_{2N})_{\mathrm{axial}}$ symmetry under the background gauge field $(A,B)$ is broken as 
\be
\calZ[(A,B)]\mapsto \calZ[(A,B)]\exp\left(-{2\im N\over 4\pi}\int B\wedge B\right). 
\label{eq:4d_anomaly}
\ee
For $N\ge 3$, this additional phase is nontrivial, and the 't~Hooft anomaly exists between the flavor symmetry $SU(N)_{\mathrm{flavor}}/(\mathbb{Z}_N)_{\mathrm{color-flavor}}$ and the discrete axial symmetry $(\mathbb{Z}_{2N})_{\mathrm{axial}}$.  
Four-dimensional QCD is believed to break the chiral symmetry spontaneously, which also breaks $\mathbb{Z}_{2N}$ axial symmetry to $\mathbb{Z}_2=\{1,(-1)^F\}$ spontaneously ($F$ is the fermion number operator), and the 't~Hooft anomaly is matched.

\subsection{Massless $\mathbb{Z}_N$-QCD and its anomaly}\label{sec:Zn_qcd_3d}

We compactify one-direction, and derive the associated three-dimensional effective theory. We fix the $SU(N)_{\mathrm{flavor}}$ holonomy as\footnote{Only if $\phi$ takes a special value, $\Omega\in SU(N)$. We can, however, perform the twist using the vector-like $U(1)$ symmetry, which does not affect the argument below. The following discussion is valid for any $\phi$. }
\be
\Omega=\mathrm{e}^{\im \phi}\mathrm{diag}[1,\omega,\omega^2,\ldots,\omega^{N-1}]. 
\label{eq:Zn_qcd_flavor_twist_matrix}
\ee
Equivalently, we introduce the boundary condition on the quark field $\Psi$ as 
\be
\Psi(\bm{x},x^4+L)=\Psi(\bm{x},x^4)\Omega. 
\ee
The extended gauge transformation eliminates the holonomy, but the quark field obeys the $\mathbb{Z}_N$-twisted boundary condition. This is called $\mathbb{Z}_N$-QCD, and we denote its partition function as $\calZ_{\Omega}$. 

Circle compactification induces $\mathbb{Z}_N$ zero-form transformation, $\Omega\mapsto \omega\Omega$, from the $\mathbb{Z}_N$ one-form symmetry (\ref{eq:Zn_one-form_QCDbifund}), but it changes the boundary condition and maps a theory $\calZ_{\Omega}$ to another theory $\calZ_{\omega \Omega}$.  
We should intertwine it with the flavor rotation $\Psi\mapsto \Psi S$, where $S$ is defined in (\ref{eq:formalism_Omega_Shift}), in order to maintain the boundary condition. This generates the $\mathbb{Z}_N$ zero-form symmetry of $\calZ_{\Omega}$, and we call this as the shift symmetry, $(\mathbb{Z}_N)_S$, which acts on local operators on $\mathbb{R}^3$ as 
\be
\Psi\mapsto \Psi S,\; \; 
\mathrm{tr}\left[\mathcal{P}\exp\im \int_{S^1}a\right]\mapsto \omega\, \mathrm{tr}\left[\mathcal{P}\exp\im \int_{S^1}a\right]. 
\ee
To obtain the three-dimensional anomaly, we gauge the shift symmetry and denote the corresponding gauge field as $B^{(1)}$. 
Because of the holonomy $\Omega$, the explicit breaking of the flavor symmetry occurs $SU(N)_\mathrm{flavor}\to U(1)^{N-1}$, and the faithful flavor symmetry is $U(1)^{N-1}/(\mathbb{Z}_N)_{\mathrm{color-flavor}}$. We introduce the $U(1)^{N-1}$ background gauge field $A_K$ and three-dimensional $\mathbb{Z}_N$ two-form gauge field $B^{(2)}$. The $\mathbb{Z}_N$-twisted partition function under these backgrounds are given by 
\be
\calZ_{\Omega}[(A_K,B^{(1)},B^{(2)})]=\calZ[(A_K+B^{(1)}+A_{\mathrm{cl}}, B^{(2)}+B^{(1)}\wedge L^{-1}\diff x^4)]. 
\ee
We thus obtain under the $(\mathbb{Z}_{2N})_{\mathrm{axial}}$ transformation as 
\be
\calZ_{\Omega}[(A_K,B^{(1)},B^{(2)})]\mapsto \calZ_{\Omega}[(A_K,B^{(1)},B^{(2)})]\exp\left(-{2\im N\over 2\pi}\int B^{(2)}\wedge B^{(1)}\right). 
\label{eq:anomaly_Zn_qcd}
\ee
Therefore, there is a mixed 't Hooft anomaly among the shift symmetry $(\mathbb{Z}_N)_S$, the flavor symmetry $U(1)^{N-1}/(\mathbb{Z}_N)_{\mathrm{color-flavor}}$, and the discrete axial symmetry $(\mathbb{Z}_{2N})_{\mathrm{axial}}$, for $N\ge 3$.

\subsection{Comparison with previous studies and Discussion}

Since $\mathbb{Z}_N$-QCD has the $\mathbb{Z}_N$ global symmetry acting on the Polyakov loop $\Phi=\mathrm{tr}\left[\mathcal{P}\exp\im \int_{S^1}a\right]$, it has been used as a tool to study the confinement-deconfinement transition of QCD by effective models~\cite{Kouno:2012zz, Sakai:2012ika, Kouno:2013zr, Kouno:2013mma}, by reliable semiclassical analysis of softly-broken $\mathcal{N}=1$ supersymmetric QCD on $\mathbb{R}^3\times S^1$~\cite{Poppitz:2013zqa}, and also by lattice simulation~\cite{Iritani:2015ara}. 
The same boundary condition is also used for different and various purposes, such as the reduction of finite-volume effect in lattice simulations, definition of new order parameters, etc.~\cite{Briceno:2013hya, Cherman:2016hcd, Liu:2016yij, Cherman:2016vpt,Cherman:2017tey}\footnote{It is also notable that the similar (but distinct) twisted boundary condition works as a nonperturbative renormalized-coupling scheme in finite volume on the lattice~\cite{Bilgici:2009nm, Itou:2010we, Itou:2012qn, Itou:2013kaa, Itou:2013faa}.}. 
We especially compare the 't Hooft anomaly (\ref{eq:anomaly_Zn_qcd}) with the lattice simulation about thermodynamic properties of $\mathbb{Z}_N$-QCD~\cite{Iritani:2015ara}. 

Simulation setup in Ref.~\cite{Iritani:2015ara} is massive $\mathbb{Z}_3$-QCD with relatively heavy quarks, $m_{\mathrm{PS}}/m_{\mathrm{V}}=0.70$, where we denote $\Psi=(u,d,s)$. 
The flavor twist matrix $\Omega$ in (\ref{eq:Zn_qcd_flavor_twist_matrix}) is taken as $\phi=\pi$. 
The first-order chiral and center phase transition is identified by observing hysteresis behavior of expectation values of chiral condensate and Polyakov loop depending on  thermalization process, even though the chiral phase transition is obscured by the fermion mass. 
At low temperature, the axial symmetry $(\mathbb{Z}_{2N})_{\mathrm{axial}}$ is spontaneously broken, and the shift symmetry $(\mathbb{Z}_N)_S$, which is called as the intertwined color-flavor symmetry, is unbroken since $\langle \Phi\rangle=0$, where three chiral condensates for $u,d,s$ give the same value, $\langle\overline{u}u\rangle=\langle\overline{d}d\rangle=\langle\overline{s}s\rangle$. 
At high temperature, the intertwined color-flavor center symmetry $(\mathbb{Z}_N)_S$ is spontaneously broken, i.e., $\langle \Phi\rangle\not=0$. 
In both cases, the 't Hooft anomaly given in (\ref{eq:anomaly_Zn_qcd}) is matched. 

What is claimed by anomaly matching is that trivial phase is denied even in the intermediate regime. In the lattice simulation~\cite{Iritani:2015ara}, the transition temperatures of first-order phase transition are almost equal for the Polyakov loop and for the chiral condensate. This is also consistent with the implication of anomaly. 

Finite density $\mathbb{Z}_3$-QCD is also studied recently in the context of effective models~\cite{Kouno:2015sja, Hirakida:2016rqd, Hirakida:2017bye}, and the new order parameter of QCD can be defined with the $\mathbb{Z}_3$ twisted boundary condition~\cite{Cherman:2017tey}. 
Extension of our technique to the case with chemical potential is an important future work, and the implication of anomaly matching will be quite interesting. 
Our systematic construction of three-dimensional anomaly of massless QCD will provide a basic guideline for that purpose. 
Indeed, the chemical potential $\mu$ is the imaginary-valued constant background of vector-like $U(1)$ gauge field, and the above discussion is not changed by the existence of $\mu$ at least in the formal level. 
If we assume no unexpected obstruction against our anomaly 
computation (\ref{eq:anomaly_Zn_qcd}) in the presence of $\mu$, we can obtain an ambitious 
conclusion that the trivial gapped phase does not appear in the phase diagram of massless $\mathbb{Z}_N$-QCD. 
This possibility will be investigated in another paper in detail~\cite{Tanizaki:2017mtm}. 

Anomaly of massless QCD with $N_c=N_f$ at finite temperature with an imaginary chemical potential is discussed in Ref.~\cite{Shimizu:2017asf}, and Roberge-Weiss phase transition~\cite{Roberge:1986mm} is reconsidered from the viewpoint of anomaly. 
The construction of anomaly for $S^1$ compactified theory is based on the same idea. It is an interesting work to reproduce their anomaly starting from the anomaly of four-dimensional massless QCD with our systematic procedure by paying attention to different symmetries and employing different twisted boundary conditions.

\section{Conclusion}\label{sec:conclusion}

We developed a systematic procedure that derives an 't Hooft anomaly of $D$-dimensional theories after circle compactification, and clarified its connection to the 't Hooft anomaly of the original $(D+1)$-dimensional theory. 
This applies to the case when the $(D+1)$-dimensional theory has no higher-form symmetries, and the appropriate choice of the twisted boundary condition is important there. 

We compute 't Hooft anomalies under circle compactification for two theories: $\mathbb{Z}_N$-twisted $\mathbb{C}P^{N-1}$ model and massless $\mathbb{Z}_N$-QCD. 
Reliable semiclassical computation is applicable for $\mathbb{Z}_N$-twisted $\mathbb{C}P^{N-1}$ model, and anomaly matching claims a consistent result. 
$\mathbb{Z}_N$-QCD has also been studied in oder to find the nature of confinement-deconfinement transition, and anomaly matching also claims a consistent result with that of numerical lattice simulation. 
Our systematic construction of anomaly via circle compactification elucidates that these behaviors are controlled by the anomaly of original theory, and trivial gapped phase is forbidden for these systems. 

It is an interesting and important task to extend our construction of anomaly starting from the original theory to the case with chemical potential. 
Quite typically, sign problem appears~\cite{Loh:1990zz, Batrouni:1992fj, Cohen:2003kd, Tanizaki:2015rda} if the chemical potential is introduced to the system, and numerical lattice simulation is still unavailable. It therefore has a great impact to claim something rigorous based on QFT to such systems. 
For example, our construction of three-dimensional anomaly for massless $\mathbb{Z}_N$-QCD seems to be valid at finite chemical potential $\mu$ at least in the formal level. If unprecedented problems do not appear, our computation of the 't Hooft anomaly ambitiously claims that there is no trivial gapped phase in the phase diagram of massless $\mathbb{Z}_N$-QCD. 

Another possible application of our technique is to study the physics of domain walls~\cite{Anber:2015kea, Sulejmanpasic:2016uwq}. 
In a certain setup of domain walls, we can use anomaly inflow to discuss properties on localized degrees of freedom living in domain walls~\cite{Gaiotto:2017yup, Komargodski:2017smk, Gaiotto:2017tne, Callan:1984sa}. 
Our construction of anomalies of circle-compactified theories claims that the physics of domain walls is also constrained  by the same anomaly of the original theory under circle compactifications with twisted boundary conditions. 

\acknowledgments
Y.~T. thanks Yuta Kikuchi for collaboration on related works. 
This work is started at the conference ``RIMS-iTHEMS International Workshop on Resurgence Theory'' at Kobe, Japan in 6-8 September 2017, and is completed during the workshop ``Resurgent Asymptotics in Physics and Mathematics'' at Kavli Institute for Theoretical Physics from October 2017. 
The authors greatly appreciate these opportunities and hospitalities of organizers. 
Y.~T. is financially supported by RIKEN special postdoctoral program. 
This work is supported in part by the Japan Society for the 
Promotion of Science (JSPS) Grant-in-Aid for Scientific Research
(KAKENHI) Grant Numbers 16K17677 (T.~M).
T.~M and N.~S are also supported by MEXT-Supported Program for the Strategic Research Foundation
at Private Universities (Keio University) ``Topological Science" (Grant No. S1511006).
Research at KITP is supported by the National Science Foundation under
Grant No. NSF PHY-1125915.

\appendix

\bibliographystyle{utphys}
\bibliography{./QFT,./QM,./lefschetz}
\end{document}